\documentclass[]{spie}  
\usepackage[dvips]{graphicx}

\newcommand\arcsec{\mbox{$^{\prime\prime}$}}%

\title{Preliminary Abundance Analysis of Galactic Bulge Main Sequence,  Subgiant, and Giant Branch Stars Observed During Microlensing with Keck/HIRES\footnote{This work was performed under the auspices of the U.S. Department of Energy,
National Nuclear Security Administration by the University of California,
Lawrence Livermore National Laboratory under contract No. W-7405-Eng-48.}}

\author{R. M. Cavallo\supit{a}, K. H. Cook\supit{a}, D. Minniti\supit{b},
        and T. Vandehei\supit{c}
\skiplinehalf
\supit{a}Lawrence Livermore National Laboratory, 7000 East Ave, L-413, Livermore, CA 94550\\
\supit{b}Universidad de Cat\'{o}lica, Chile\\ 
\supit{c}University of California San Diego, San Diego, CA
}

\authorinfo{Further author information: (Send correspondence to R.M.C.)\\
R.M.C.: E-mail: rcavallo@llnl.gov, Telephone 1 925 422 3277}

\begin{document} 
  \maketitle 

\begin{abstract}

We present an abundance analysis of six main sequence turnoff,
subgiant, and giant branch stars toward the Galactic bulge that were observed with
Keck/HIRES during microlensing events.   This is an early look at the first detailed
chemical analysis of main sequence stars in the Galactic bulge.
Lensing events allow the effective aperture of Keck to be increased beyond
its current dimensions; although, some events still stretched its spectroscopic 
capabilities.  Future large telescopes with high
resolution and high throughput spectrometers will allow the study of abundances
in distant stellar populations and in less evolved stars with greater ease.
\end{abstract}

\keywords{MACHO Project, Bulge, Abundances, Galactic Chemical Evolution}

\section{INTRODUCTION}
\label{sect:intro}  

The distance to the Galactic bulge has until recently kept the prospects of
 observing directly the chemical properties of its constituent stars just
 past the edge of possibility.
Even now, with large-aperture telescopes such as the Kecks, the VLT,
 the HET, and the Gemini twins among others,
 only the brightest giants in the Bulge have been observable~\cite{RM2000}.
A novel approach employed by the MACHO microlensing team has taken advantage of
 the temporary brightening of the source star during the lensing 
 event to effectively decrease that once prohibitive distance to the Galactic center
 and allow for the observation of less evolved stars.
For example, Minniti et al.~\cite{Minniti} first published results in 1997 on the presence
 of a Li I feature observed in a hot ($T_{\rm eff}~=~6000$~K) main sequence turn-off star.
In this paper we discuss the full analysis of the that spectrum along with the
 spectra of five other stars observed during lensing events between 1997 and
 1999.
The results to date are still preliminary; however, they offer a glimpse into
 the diverse chemical evolution scenarios that can be found in and around the
 center of the Milky Way.
Future telescopes with effective apertures in the 30-m or more range will
 make such observations commonplace rather than rare.

\section{OBSERVATIONS}
\label{sect:obs}

Between August 1997 and July 1999 we observed with the
 Keck I telescope and the HIRES echelle spectrometer 18 microlensing events that
 occurred when a small compact object (e.g., a normal star) crossed the line
 of sight between earth and a Bulge member, causing a temporary increase in
 the brightness of the latter by gravitational lensing.
Of these, only six spectra had a sufficiently high signal-to-noise ratio
 to undergo a complete analysis that allowed us to determine: 1) radial velocities;
 2) atmospheric parameters; and 3) chemical abundances.
These six stars are described in Table~\ref{tab:stars}, where they are identified by their
 lensing event number and their MACHO id in the first two columns, 
 their unlensed magnitude and color in the next
 two columns, the epoch of their observation and the exposure time in the fifth and
 sixth columns, and an estimate of the signal-to-noise ratio in the last column.
The ($V~-~R)_{0}$ colors are corrected based on recent extinction maps by Popowski,
 Cook, and Becker~\cite{Popowski}.
We note that the events 97-BLG-45 and 97-BLG-47 were indeed the same star identified
 separately in overlapping MACHO fields.
The event 98-BLG-6 was observed spectroscopically during two separate epochs in 
 1998, with the spectra being combined after ensuring no variation in the relative quality of
 each set of observations.
The signal-to-noise ratio of this event represents that of the combined (16,200 s)
 spectra.
Information regarding these and other microlensing events can be found on the MACHO
 project home page at {\tt http://www.macho.mcmaster.ca/}.

\begin{table}[htb]
\caption{Program star data}
\label{tab:stars}
\begin{center}       
\begin{tabular}{|l|c|c|c|c|r|c|}
\hline
\rule[-1ex]{0pt}{3.5ex} Lensing Event & F.T.S & $V$ & $(V-R)_0$ & Epoch & t$_{\rm exp}$ (s) & S/N \\
\hline
\rule[-1ex]{0pt}{3.5ex} 97-BLG-45/47  & 142.27650.6057/136.27650.2370 &19.8 & 0.42 & 28 Aug 1997  & 10800 & 40  \\
\hline
\rule[-1ex]{0pt}{3.5ex} 97-BLG-56     & 403.47671.57 & 16.1 & 0.85 & 27 Aug 1997  &   450 & 80  \\
\hline
\rule[-1ex]{0pt}{3.5ex} 98-BLG-6      & 402.48103.1719 & 19.5 & 0.59 & 20 June 1998 & 10800 & 60  \\
\hline
\rule[-1ex]{0pt}{3.5ex}               &  &     &       & 18 Aug 1998  &  5400 &     \\
\hline
\rule[-1ex]{0pt}{3.5ex} 99-BLG-1      & 121.22423.1032 &18.8 & 0.58 &  6 July 1999 &  7200 & 65  \\
\hline
\rule[-1ex]{0pt}{3.5ex} 99-BLG-8      & 403.47849.756  &15.8 & 0.82 &  6 July 1999 &  1800 & 130 \\
\hline
\rule[-1ex]{0pt}{3.5ex} 99-BLG-22     & 109.20893.3423 & 19.6 & 0.79 &  7 July 1999 &  2700 & 90  \\
\hline
\end{tabular}
\end{center}
\end{table}

Our setup provided us with wavelength coverage from 4780~{\AA} to 7150~{\AA}
 with gaps between the orders that increased with wavelength.
This allowed for the observation of many atomic absorption lines; however,
 it missed the important [O~I] feature at ${\lambda}$6300.
We binned the data $2~{\times}~2$ in order to increase signal and decrease
 readout times.
Combined with typical seeing of 1{\arcsec}, this produced a resolution of R~${\sim}$~29,000;
 sufficient for doing abundances work.

\begin{figure}[htb]
   \begin{center}
   \begin{tabular}{c}
   \includegraphics[height=7cm]{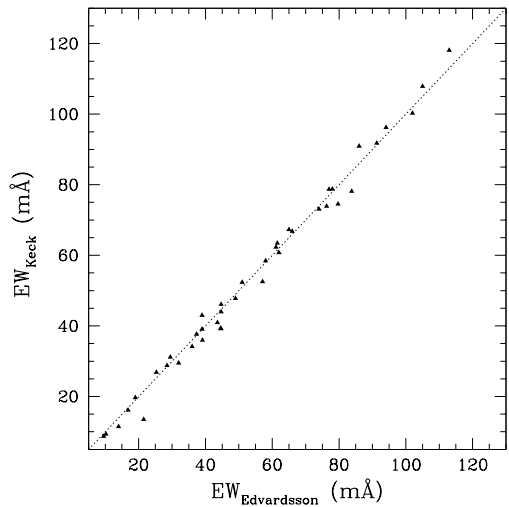}
   \end{tabular}
   \end{center}
   \caption[example]
   { \label{fig:ewcomp}
  Equivalent width comparison between our data and Edvardsson et al.~{\cite{Edvardsson}
 for the star HR~5019.  The dotted line represents perfect agreement.}
}
   \end{figure}

The data were reduced using the batch mode of MAKEE, written by T.~Barlow
 and modified by him to accommodate our $2~{\times}~2$ binning format.
MAKEE takes raw images as input and, after correcting for bias and flat fielding,
 extracts and wavelength calibrates the spectra.
We then converted the MAKEE fits output files into an IRAF readable format using the
 MAKEE routine {\tt linear}.
Once the spectra were extracted we used the spectrum of
 a rapidly rotating hot star to correct for telluric lines by dividing the 
 object spectrum by that of the rapid rotator, with the two spectra scaled and
 shifted by hand in the IRAF routine {\tt telluric}.
This removed most of the blaze function, allowing us to fit a low order polynomial
 to the data to produce flat, normalized spectra.
We then measured the equivalent widths of lines by fitting gaussians with the
 IRAF routine {\tt splot}.
To test the quality of the data reduction, we reduced the spectra of many comparison
 stars that were taken with our program stars and that have equivalent
 widths published in the literature (e.g., Edvardsson et al.~\cite{Edvardsson}).
We found the agreement between our equivalent widths and the published data to
 be generally excellent, as shown in Fig.~\ref{fig:ewcomp} for a sample star.

\section{ANALYSIS}
\label{sect:analysis}

We began our analysis by determining heliocentric-corrected radial velocities.
To do this we used the IRAF routine {\tt rvidlines} and the iron line list of
 Nave et al.~\cite{Nave}, from which we selected approximately 450 Fe~I lines 
 with accurate wavelengths that fell within our spectral coverage.
During each epoch of observations we observed comparison stars with well determined radial
 velocities, and our analysis of the data showed excellent agreement between the radial velocities
 determined for these and their published values.
The final column of Table~\ref{tab:params} lists the radial velocities of our program
 stars.

Using the stellar atmosphere code MARCS (Gustafsson et al.~\cite{MARCS})
 and the LTE abundance analysis code MOOG (Sneden~\cite{MOOG}), 
 we determined model atmosphere parameters for our program stars 
 by using the Fe~I and Fe~II lines to achieve: 1) no trend in abundance versus
 excitation potential to determine T$_{\rm eff}$;
 2) no trend in abundance versus reduced width to determine microturbulence (${\xi}$);
 and 3) agreement in abundances from neutral and ionized atoms to determine log~$g$.
We used ${\sim}$~49 Fe~I lines from Nave et al.~\cite{Nave} and ${\sim}$~14 Fe~II
 lines from Bi\'{e}mont et al.~\cite{FeII}, with the latter's log$gf$ values corrected to 
 log${\epsilon}$(Fe)~=~7.52 (a scale relative to hydrogen
  where log(H) is taken to be 12.00).
The estimated uncertainties are ${\pm}$200~K in $T_{\rm eff}$, ${\pm}$0.5 dex in log$~g$,
 and ${\pm}$0.2~km~s$^{-1}$ in ${\xi}$.
The derived atmospheric parameters are listed in Table~\ref{tab:params}.

We could have used the $V-R$ colors and the scales of Alonso et al.~\cite{Alonso} or
 Houdashelt, Bell, and Sweigart~\cite{HBS} plus some assumptions about the
 distance modulus of each star to determine atmospheric parameters, but the
 uncertainties in the reddening are too large for this approach to be any
 more certain than using the iron lines.

Column 5 of Table~\ref{tab:params} lists the derived MK classifications, 
 which were found using line-depth ratios (e.g., Strassmeier and Fekel~\cite{SF} or
 Gray and Johanson~\cite{gray}) and 
 are not from the derived atmospheric parameters; however, it should be noted that
 in general the agreement between the physical parameters and the classification is
 quite good.

\begin{table}[htb]
\caption{Derived stellar parameters}
\label{tab:params}
\begin{center}       
\begin{tabular}{|l|c|c|c|c|l|}
\hline
\rule[-1ex]{0pt}{3.5ex} Lensing Event & $T_{\rm eff}$ (K) & log~$g$ (cm s$^{-1}$) & ${\xi}$ (km~s$^{-1}$) & MK class & V$_{\rm rad}$ (km~s$^{-1}$) \\
\hline
\rule[-1ex]{0pt}{3.5ex} 97-BLG-45/47  &  6000  &  4.30 &  2.48 & G5 V     &  $+59.98~{\pm}~2.24$ \\
\hline
\rule[-1ex]{0pt}{3.5ex} 97-BLG-56     &  4480  &  1.40 &  1.84 & G8 III   &  $+177.0~{\pm}~1.9$   \\
\hline
\rule[-1ex]{0pt}{3.5ex} 98-BLG-6      &  5200  &  3.50 &  2.28 & G8 IV    &  $-64.24~{\pm}~1.68$ \\
\hline
\rule[-1ex]{0pt}{3.5ex} 99-BLG-1      &  5600  &  4.65 &  3.30 & G2 V     &  $+64.10~{\pm}~1.78$ \\
\hline
\rule[-1ex]{0pt}{3.5ex} 99-BLG-8      &  4150  &  0.95 &  1.91 & K2 III   &  $+194.9~{\pm}~1.7$  \\
\hline
\rule[-1ex]{0pt}{3.5ex} 99-BLG-22     &  5885  &  4.30 &  1.52 & F0 V     &  $+37.35~{\pm}~2.06$ \\
\hline
\end{tabular}
\end{center}
\end{table}

\section{RESULTS}
\label{sect:results}

In the Tables~3-6 we present an early look into the abundances of our program
 stars, with the caveat that some numbers are subject to change;
 i.e., elements that suffer from heavy hyperfine splitting (hfs) and/or isotopic
 shifts will likely be revised in future work~\cite{us}.
Currently, we correct for such effects when possible by correcting one or
 two lines with the atomic data of Prochaska~\cite{Prochaska} and applying an 
 estimated correction to the total abundance.
Abundances listed with a subscript (e.g., [Sc/Fe]$_{\rm I}$ and [Sc/Fe]$_{\rm II}$)
 are derived from both the neutral lines (I) and the singly ionized lines (II),
 and are listed separately.
In all other cases the abundances come from lines of either ionization state.

Tables~3-6 do not list uncertainties for abundances other than iron;
 typically errors from line-to-line scatter for a given element are ${\sim}$~0.2 dex but
 can vary quite a bit.
This is especially true for silicon where the rms is more like 0.5 dex; however,
 [Si/Fe] is determined from lines with excitation potential around 6 e.V., while
 atmospheric parameters came from Fe~I lines with ${\chi}~<~5$~e.V..
On the other hand [Al/Fe] and [Na/Fe] often showed good agreement between
 the abundances derived from different lines (${\sigma}~<~0.1$ dex).

We discuss the quality of the data for each star individually in Section~\ref{sect:summary}.

\begin{table}[htb]
\caption{Derived abundances}
\label{tab:abuns01}
\begin{center}
\begin{tabular}{|l|c|c|c|c|c|c|c|c|}
\hline
\rule[-1ex]{0pt}{3.5ex} Lensing Event & [Fe/H] & [Na/Fe] & [Mg/Fe] & [Al/Fe] & [Si/Fe] & [Ca/Fe] & [Sc/Fe]$_{\rm I}$ & [Sc/Fe]$_{\rm II}$ \\
\hline
\rule[-1ex]{0pt}{3.5ex} 97-BLG-45/47  &  $-0.28~{\pm}~0.22$ & $+0.42$ & $+0.33$ & $+0.35$ & $+0.23$ & $+0.07$ & ... & $-0.96$ \\
\hline
\rule[-1ex]{0pt}{3.5ex} 97-BLG-56     &  $-1.13~{\pm}~0.20$ & $-0.03$ & $+0.79$ & $+0.15$ & $+0.30$ & $+0.24$ & $+0.65$ & ...\\
\hline
\rule[-1ex]{0pt}{3.5ex} 98-BLG-6     &  $-0.22~{\pm}~0.18$ & $+0.30$ & $+0.29$ & $+0.10$ & $+0.28$ & $-0.06$ & $+0.31$ & ... \\
\hline
\rule[-1ex]{0pt}{3.5ex} 99-BLG-1     &  $-0.01~{\pm}~0.28$ & $+0.67$ & $+0.81$ & $+0.67$ & $+0.33$ & $+0.13$ & $+1.36$ & $-0.12$ \\
\hline
\rule[-1ex]{0pt}{3.5ex} 99-BLG-8     &  $-1.12~{\pm}~0.20$ & $+0.21$ & $+0.62$ & $+0.40$ & $+0.42$ & $+0.09$ & $+0.63$ & $+1.14$ \\
\hline
\rule[-1ex]{0pt}{3.5ex} 99-BLG-22     &  $-0.35~{\pm}~0.20$ & $+0.12$ & $+0.59$ & $+0.33$ & $+0.34$ & $+0.20$ & ... & $-0.45$ \\
\hline
\end{tabular}
\end{center}
\end{table}

\begin{table}[htb]
\caption{Derived abundances (continued)}
\label{tab:abuns02}
\begin{center}
\begin{tabular}{|l|c|c|c|c|c|c|c|c|}
\hline
\rule[-1ex]{0pt}{3.5ex} Lensing Event & [Ti/Fe]$_{\rm I}$ & [Ti/Fe]$_{\rm II}$ & [V/Fe]$_{\rm I}$ & [V/Fe]$_{\rm II}$ & [Cr/Fe]$_{\rm I}$ & [Cr/Fe]$_{\rm II}$ & [Mn/Fe] & [Co/Fe] \\
\hline
\rule[-1ex]{0pt}{3.5ex} 97-BLG-45/47  & $+0.92$ & $+0.50$ & $+0.52$ & $+1.23$ & $+0.45$ & ... & $-0.25$ & $+0.61$ \\
\hline
\rule[-1ex]{0pt}{3.5ex} 97-BLG-56     & $+0.67$ & $+0.83$ & $+0.61$ & ... & $+0.45$ & $+0.14$ & $-0.25$ & $-0.01$ \\
\hline
\rule[-1ex]{0pt}{3.5ex} 98-BLG-6     & $+0.50$ & $+0.43$ & $+0.54$ & ... & $+0.08$ & ... & $+0.08$ & $+0.00$ \\
\hline
\rule[-1ex]{0pt}{3.5ex} 99-BLG-1     & $+1.19$ & $+0.39$ & $+1.33$ & $+1.11$ & $+0.55$ & ... & $+0.81$ & $+0.46$ \\
\hline
\rule[-1ex]{0pt}{3.5ex} 99-BLG-8     & $+0.68$ & $+0.34$ & $+0.17$ & $+0.48$ & $+0.35$ & $+0.75$ & $-0.28$ & $-0.15$ \\
\hline
\rule[-1ex]{0pt}{3.5ex} 99-BLG-22     & $+0.70$ & $+0.47$ & $+0.52$ & $+1.47$ & $+0.01$ & ... & $-0.27$ & $+0.60$ \\
\hline
\end{tabular}
\end{center}
\end{table}

\begin{table}[htb]
\caption{Derived abundances (continued)}
\label{tab:abuns03}
\begin{center}
\begin{tabular}{|l|c|c|c|c|c|c|c|c|c|}
\hline
\rule[-1ex]{0pt}{3.5ex} Lensing Event & [Ni/Fe] & [Zn/Fe] & [Y/Fe]$_{\rm I}$ & [Y/Fe]$_{\rm II}$ & [Zr/Fe] & [Nb/Fe] & [Mo/Fe] & [Ru/Fe] & [Ba/Fe]  \\
\hline
\rule[-1ex]{0pt}{3.5ex} 97-BLG-45/47  & $+0.28$ & $+0.29$ & ...     & $+0.35$ & ... & ...     & ...     & ...  & $+0.39$ \\
\hline
\rule[-1ex]{0pt}{3.5ex} 97-BLG-56     & $+0.25$ & ...     & $+0.84$ & $+0.42$ & $+1.14$ & ... & $+1.14$ & $+0.92$ & $+0.57$ \\
\hline
\rule[-1ex]{0pt}{3.5ex} 98-BLG-6     & $+0.19$ & $+0.48$ & $+0.92$ & $+0.91$ & $+0.83$ & ... & ... & ... & $+0.66$ \\
\hline
\rule[-1ex]{0pt}{3.5ex} 99-BLG-1     & $+0.31$ & $-0.22$ & $+1.43$ & $+0.96$ & $+1.76$ & ... & $+1.31$ & $+1.94$ & $+0.01$ \\
\hline
\rule[-1ex]{0pt}{3.5ex} 99-BLG-8     & $+0.12$ & $-0.03$ & $+0.76$ & $+0.61$ & $+1.15$ & $+1.54$ & $+0.92$ & $+1.11$ & $+0.68$ \\
\hline
\rule[-1ex]{0pt}{3.5ex} 99-BLG-22     & $+0.23$ & $+0.03$ & ... & $+0.05$ & ... & ... & ... & ... & $+0.26$ \\
\hline
\end{tabular}
\end{center}
\end{table}

\begin{table}[htb]
\caption{Derived abundances (continued)}
\label{tab:abuns04}
\begin{center}
\begin{tabular}{|l|c|c|c|c|c|}
\hline
\rule[-1ex]{0pt}{3.5ex} Lensing Event & [La/Fe] & [Ce/Fe] & [Pr/Fe] & [Nd/Fe] & [Eu/Fe]\\
\hline
\rule[-1ex]{0pt}{3.5ex} 97-BLG-45/47  & $+0.97$ & ... & ... & $+1.72$ & ...\\
\hline
\rule[-1ex]{0pt}{3.5ex} 97-BLG-56     & $+0.81$ & $+1.27$ & $+0.90$ & $+0.99$ & $+0.53$ \\
\hline
\rule[-1ex]{0pt}{3.5ex} 98-BLG-6     & $+0.99$ & $+1.10$ & $+0.83$ & $+0.92$ & $+1.09$ \\
\hline
\rule[-1ex]{0pt}{3.5ex} 99-BLG-1     & $+1.47$ & $+1.55$ & ...     & $+1.31$ & $+1.46$ \\
\hline
\rule[-1ex]{0pt}{3.5ex} 99-BLG-8     & $+0.91$ & $+1.11$ & $+0.94$ & $+0.85$ & $+1.34$ \\
\hline
\rule[-1ex]{0pt}{3.5ex} 99-BLG-22     & ...     & $+2.39$ & $+0.89$ & $+0.89$ & ...     \\
\hline
\end{tabular}
\end{center}
\end{table}

\section{Summary of Results} 
\label{sect:summary}
\subsection{97-BLG-56 and 99-BLG-8}

We begin our discussion of the program stars with the cool giants.  
Both spectra have excellent signal-to-noise ratios, which comes
 as no surprise since these stars are already quite bright and
 being aided by lensing makes them that much brighter.
They both contain the lowest [Fe/H] in the sample, raising
 the possibility of systematic errors in their temperatures
 affecting the iron abundance determinations.
We doubt this is the case since our reduction techniques have been
 compared favorably with our so-called ``standard stars'' and our model
 parameters match well with the MK classifications derived
 via line-depth ratios.
The radial velocities of these two stars are the highest in
 the sample, giving rise to speculation that they could
 be members of the Sagittarius dwarf spheroidal; however,
 using the spectral type to estimate the absolute magnitudes
 and some assumptions about reddening~\cite{Popowski}, we derive distances
 to these stars of roughly 3.3~-~3.9 kpc.
Compared with the Sagittarius dwarf spheroidal, which has a distance
 of ${\sim}~28$~ kpc~\cite{SL95}, these two stars clearly belong to a
 separate stellar population, even when accounting for errors in reddening
 or spectral type.

The abundance trends in both stars are consistent: the low-Z (Z~$<~40$)
 elements are enhanced by about 0.5 dex while the high-Z elements
 are up by around 1 dex, indicating enhancements of $s$-process
 from possible AGB stars.
The elements with estimated hfs corrections are V, Mn, and Co for both
 stars, leaving the abundances from the high-Z elements perhaps overestimated;
 although, not all of these require hfs corrections.

\subsection{98-BLG-6}

98-BLG-6 is roughly solar metallicity, particularly when the error bars are
 taken into account.
The luminosity class is estimated from log$~g$ as either a main sequence turnoff
 star or a subgiant.
The gravity does seem to be well determined given the excellent agreement between
 the abundances from the neutral and ionized atoms for the various species when
 available.
The effective temperature/log$~g$ combination along with the spectral type from
 line ratios is consistent with this star being a subgiant (see, e.g., Th\'{e}venin
 and Jasniewicz~\cite{G8IV}), making 98-BLG-6 the only one in this study.
The radial velocity for this star differs from the previous two by 250~km~s$^{-1}$
 and is the only one of the six that is negative.
Estimated hfs corrections were done for V, Mn, and Co with corrections amounting to
 $-0.65$ dex for [Mn/Fe] and $-0.45$ dex for [Co/Fe].
Given these large corrections for the iron-peak elements, the high-Z elemental
 abundances should be viewed with caution.
Finally, it should be noted that we are able to determine the abundances of 22 elements 
 in a $V~=~19.5$ star only because of microlensing.
Such detailed calculations are the exception at this magnitude, but should become
 trivial with a new age of 30-m and larger class telescopes, provided they
 are outfitted with the proper instrumentation and can function in the visible
 bands.

\subsection{99-BLG-1}

According to our spectral typing, 99-BLG-1 is a solar analog in the Galactic bulge.
Our derived physical parameters are also consistent with solar-like parameters
 within their errors.
Some lines from Sc~II, V~I, Mn~I, and Co~I have been corrected for hfs effects.
Both [Sc/Fe]$_{\rm II}$ and [V/Fe]$_{\rm II}$ change very little as a result, which
 is surprising given the large overabundance of [V/Fe]$_{\rm II}$; however, there
 are 18 other V~II lines that have yet to be corrected and it remains to be seen
 whether the one line we corrected at ${\lambda}$5671 is anomalous or is actually
 indicative of large V enhancements.
The high-Z elements as a whole show the largest enhancements of all the stars.
Ruthenium is particularly enhanced; however, the abundance comes from only
 one line at ${\lambda}$5636 and this element does have seven stable isotopes with a
 rather flat distribution in relative abundance in nature among them.
Barium is the one anomalous high-Z element in this star with a scaled solar ratio;
 this despite it's sister element, La, being enhanced by a factor of 30 relative
 to the sun.
No hfs corrections have been applied to either of these elements.
On the other hand, europium is enhanced and the high Eu/Ba ratio would indicate
 heavy $r$-process influences.
Further analysis of this star should yield an interesting chemical history.
 
\subsection{99-BLG-22}

This star is close to being a solar analog in temperature and gravity; however,
 the derived spectral type appears to make it too hot.
The signal-to-noise ratio is quite good for this star (especially considering
 it has an unlensed $V$ magnitude of 19.6), giving us confidence in our spectral 
 analysis.
Relatively little information is available for the high-Z elements, but once 
 again, [Ba/Fe] is near solar while the heavier elements remain high.
[Sc/Fe]$_{\rm II}$, [V/Fe]$_{\rm I}$, and [Mn/Fe]$_{\rm I}$ have estimated corrections
 for hfs effects.
Despite the quality of the spectrum, abundances of Zn, Y, Ce, Pr, and Nd come from only
 one or two lines each, which is atypical compared with the brighter stars.

\subsection{97-BLG-45/47}

We conclude our summary with a discussion of 97-BLG-45/47.
Like 99-BLG-22, the derived stellar parameters are at odds with the spectral
 type, and given the poor signal-to-noise ratio in the spectrum
 it is difficult to decide on a preferred set of parameters.
The choices we adopted in Table~\ref{tab:params} are found through
 spectroscopic means and are consistent with the parameters listed in
 Minniti et al.~\cite{Minniti}, making the star a G0 dwarf.
97-BLG-45/47 is slightly underabundant in iron and shows a range of scatter 
 in the remaining elements.
[Nd/Fe] is particularly high, but the results do come from five lines and show
 only 0.34 dex scatter.
The results for [Sc/Fe]$_{\rm II}$, [V/Fe]$_{\rm I}$, and [Mn/Fe]$_{\rm I}$
 do include estimated hfs corrections.

\begin{figure}
   \begin{center}
   \begin{tabular}{c}
   \includegraphics[height=7cm]{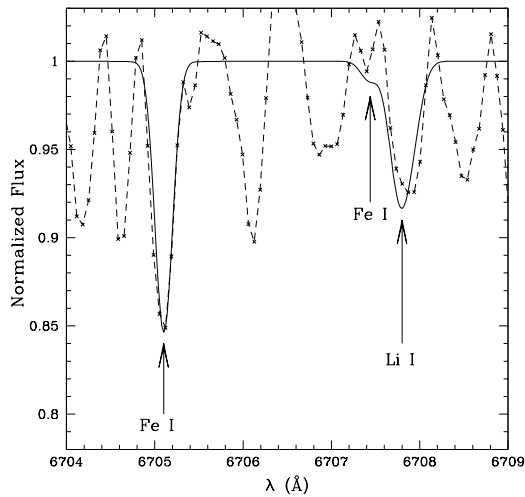}
   \end{tabular}
   \end{center}
   \caption[example] 
   { \label{fig:97blg45.li} 
  The region around the Li I doublet at ${\lambda}$6708 showing the spectrum of
   97-BLG-45/47 (x's and dotted line) and the synthetic spectrum with 
   log${\epsilon}$(Li)~=~2.25 and log${\epsilon}$(Fe)~=~7.42 (solid line).
}
   \end{figure}

We revisited the ${\lambda}$6708 Li I feature in the spectrum of 97-BLG-45/47 as
 first studied by Minniti et al.~\cite{Minniti}.
They measured an equivalent width of 47~m{\AA} and estimated an abundance of 
 log${\epsilon}$(Li)~=~2.25 from published tables.
We measure the equivalent width to be 37.3~m{\AA}, which when run in a full
 synthesis using the line list from Ford et al.~\cite{Ford}, results in 
 log${\epsilon}$(Li)~=~2.25; i.e., good agreement with the earlier work.
A question, however, arises as to the validity of the detection, given the noise in
 the spectrum and the fact that the center of the measured line is 0.12 {\AA} to the
 red of the strongest feature in the Li doublet, while the center of the Fe~I line at 
 ${\lambda}$6705 is well matched in the data, as can be seen in 
 Fig.~\ref{fig:97blg45.li}.

Finally, we note that our radial velocity is 30 km~s$^{-1}$ lower than that listed
 in Minniti et al.~\cite{Minniti} for uncertain reasons.
Given the excellent agreement we obtained between the radial velocities of
 our ``standards'' observed on the same nights as 97-BLG-45/47 and their published
 values, we are quite confident in our determination listed in Table~\ref{tab:params}.

\section{Conclusions}

Microlensing has allowed us to derive a reasonably large and detailed set of
 abundances for a group of stars that have heretofore been beyond our capabilities.
While some work on the details remains to be done before conclusions can be
 firmly drawn, we have been able to study abundance trends in stars as diverse
 as solar analogs and metal-poor red giants.
Our early results show a trend of overabundance in nearly all the elements regardless
 of metallicity or spectral class.
The two giants in the study also have the lowest metallicities and the highest
 radial velocities, and they are easily ruled out as members of the Sagittarius
 dwarf spheroidal.
We have also re-analyzed the Li abundance in the star 97-BLG-45 and have found
 that we cannot conclusively rule for or against the presence of Li.
All the spectra show some hint of Li ${\lambda}$6708 and we will examine these in
 kind to determine abundances or upper limits.

Our results are likely to change for some of the elements where hfs or
 isotopic corrections are still necessary.
When possible we corrected for hfs effects by correcting one line and
 applying a shift to the results from all the lines.
Sometimes the shift was large ($>~0.5$ dex) and sometimes it was
 nonexistent.

Thanks to a helping hand from Nature, we were able to get a glimpse at the
 functioning of Keck as a much larger telescope than its current 10-m diameter.
Future 30-m and larger class telescopes will make such observations
 commonplace and allow us to learn truly the diverse histories of the
 Galactic bulge stellar population.

\acknowledgments     
We wish to thank Stefan Keller for helping us classify our stars, Piotr Popowski 
 for providing us with extinction maps of the MACHO fields, and Tom Barlow for modifying
 his MAKEE routine to accept our data format and answering our many questions.
This research has made use of the SIMBAD database, operated at CDS, Strasbourg, France.


\end{document}